\documentclass{emulateapj}
\slugcomment{{\sc Accepted to ApJ:} June 2, 2015}

\usepackage{amsmath}
\usepackage{graphicx}
\usepackage{multirow}
\usepackage{color}
\usepackage{natbib}
 at 10truept

\def\P{{\cal P}}
\def\D{{\cal D}}

\begin{document}

\title{The mass distribution function of planets}

\author{Renu Malhotra}
\affil{Lunar and Planetary Laboratory\\ The University of Arizona\\ Tucson, AZ 85721, USA.}
\email{renu@lpl.arizona.edu}

\begin{abstract}
The distribution of orbital period ratios of adjacent planets in extra-solar planetary systems discovered by the {\it Kepler} space telescope exhibits a peak near $\sim1.5$--$2$, a long tail of larger period ratios, and a steep drop-off in the number of systems with period ratios below $\sim1.5$.  We find from this data that the dimensionless orbital separations have an approximately log-normal distribution.  
Using  Hill's criterion for the dynamical stability of two planets, we find an upper bound on planet masses such that the most common planet mass does not exceed $10^{-3.2}m_*$, or about two-thirds Jupiter mass for solar mass stars.  
%%%****
Assuming that the mass ratio and the dynamical separation (orbital spacings in units of mutual Hill radius) of adjacent planets are independent random variates, and adopting empirical distributions for these, we use Hill's criterion in a statistical way to estimate the planet mass distribution function from the observed distribution of orbital separations.  We find that the planet mass function is peaked in logarithm of mass, with a peak value and standard deviation of $\log m/M_\oplus$ of $\sim(0.6-1.0)$ and $\sim(1.1-1.2)$, respectively. 
%%%***
\end{abstract}

\section{Introduction}

%%%%%%%%%%%%%%%%%%%%%%%%%%%%%%%%%%%%%%%
%%%%%% Figure-fig1 %%%%%%%%%%%%%%%%%%%%%%%%%%
\begin{figure}[b]
\centering
\includegraphics[scale=0.35,angle=270]{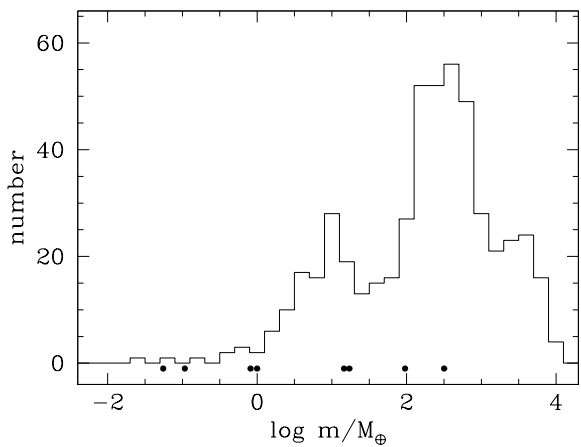}
\caption{The distribution of log-mass of confirmed exoplanets with measured masses (data from (http://exoplanetarchive.ipac.caltech.edu/, retrieved on September 16, 2014). The black points indicate the masses of the solar system planets.  Note that this is a semi-log plot.}
\label{f:fig1}
\end{figure}
%%%%%%%%%%%%%%%%%%%%%%%%%%%%%%%%%%%%%%%%%%
%%%%%%%%%%%%%%%%%%%%%%%%%%%%%%%%%%%%%%%%%%

What is the mass distribution function of planets in the Universe?  In the past two decades, more than 1700 %1743
exoplanets have been discovered; of these, 502 have measured masses (Exoplanet Archive, http://exoplanetarchive.ipac.caltech.edu/, as of September 16, 2014).  The distribution of these planet masses is shown in Fig.~\ref{f:fig1}. 
We observe two local peaks in this apparent distribution, near $\sim1~M_J$ and near $\sim10~M_\oplus$:  Jupiter-mass and ``super'' Earth-mass planets are common in the discovered population of planets. %; the $\sim1M_J$ peak is dominated by planets discovered by the radial velocity method, whereas the $\sim10~M_\oplus$ peak is dominated by planets discovered by the transit method by the {\it Kepler} mission.  
However, this apparent mass distribution suffers from many difficult-to-quantify selection biases, so we must exercise great caution in interpreting these features.
At the very least, they are of unknown significance for the intrinsic distribution of planet masses.  

In some contrast with the less--than--$30\%$ fraction of all exoplanets for which we have measured masses, the orbital periods of nearly 100\% of exoplanets are quite well determined -- indeed, the periodicity of the orbital motion of planets is predominantly how they are discovered. The {\it Kepler} mission, currently the largest systematic exoplanet survey~\citep{Borucki:2011}, has provided a wealth of data on planets and planetary systems in the Galaxy.  A large subset, about 65\%, %(1134/1743)
of all confirmed exoplanets are found in planetary systems harboring two or more planets.  
Significantly, several studies of the {\it Kepler} data on multiple-planet systems have concluded that planetary systems are coplanar to within a few degrees~\citep{Lissauer:2011, Tremaine:2012, Fang:2012b, Johansen:2012, Figueira:2012, Fabrycky:2014}, and that they are likely closely-packed~\citep{Fang:2013}.

Here we leverage the {\it Kepler} data on orbital periods in multiple planet systems, together with theoretical understanding of the long term dynamical stability of coplanar planetary systems, to estimate the planet mass distribution function.
%%%***
First, we use the observational data on orbital periods to compute the distribution of orbital spacings in systems of $N\ge2$ planets.  We then reason that it must be possible to deduce planetary masses from the observational data of orbital period ratios if orbital spacings are determined by long term dynamical stability.  Numerical studies have shown that the relationship between orbital spacing and planet masses is necessarily statistical in nature because multiple planet systems exhibit chaotic dynamics.  We adopt the ansatz that the orbital spacing measured in units of the mutual Hill radius ---the so-called ``dynamical separation''--- is a random variate, and we adopt an empirical distribution for this parameter.  This leads us to estimates of the distribution of the total mass of adjacent planets relative to the stellar host mass.  We then consider two limiting cases, that adjacent planets have a random mass ratio or that they tend to be similar to each other in mass. Finally, we convolve with the observed distribution of stellar host masses to convert from planet-to-star mass ratios to planet masses, to calculate the planet mass distribution function.   

Our approach assumes that dynamical separations, total mass and mass ratios of adjacent planets are independent random variates, and we neglect any corelations with stellar host mass and the age of the system.  With these simplifications, we arrive at a theoretical estimate of the planet mass function based only on the observational data of the orbital periods of exoplanets and the masses of their stellar hosts.  We also make independent estimates of the planet mass function based on empirical mass-radius relations and observational data of planet radii, and we compare these with the dynamical stability--based estimate.

The true mass distribution function will of course be increasingly better determined with ongoing observational efforts to measure the masses of a large population of extra-solar planets by means of complementary techniques (transits, radial velocities, astrometry, etc.).  Our simple theoretical prediction for the planet mass distribution function may be useful for the interpretation of such forthcoming observations.  Our work may also be useful for the planning and interpretation of numerical studies of the dynamical stability of planetary systems.  Knowledge of the planet mass function is important for understanding the physics of planet formation in different mass regimes, as well as for assessing the abundance of planets like our own home planet.
%%%***

\section{Orbital spacings}

\noindent
We will use the following notation: $m_*$ is the stellar mass; the planets' masses, orbital semimajor axes and orbital periods are $m_i, a_i$ and $T_i$, with $i=1,2,...$ in order of increasing distance from the host star. 
We define the mass ratios,
\begin{eqnarray}
\mu_i &=& {m_i\over m_*}, \label{e:mui}\\
\tilde{\mu_i} &=& {m_i+m_{i+1}\over m_*}, \label{e:muitilde}\\
\gamma_i &=& {\min\{m_i,m_{i+1}\}\over \max\{m_i,m_{i+1}\}}. \label{e:gamma}
\end{eqnarray}
Note that $0<\mu_i<\tilde\mu_i\ll 1$ and $0<\gamma_i\le 1$. The masses of the two nearest-neighbor planets are given by $\gamma_i(1+\gamma_i)^{-1} \tilde\mu_i m_*$ and $(1+\gamma_i)^{-1}\tilde\mu_i m_*$.

\begin{figure}[t]
\centering
\vskip-0.2truein\includegraphics[scale=0.35,angle=270]{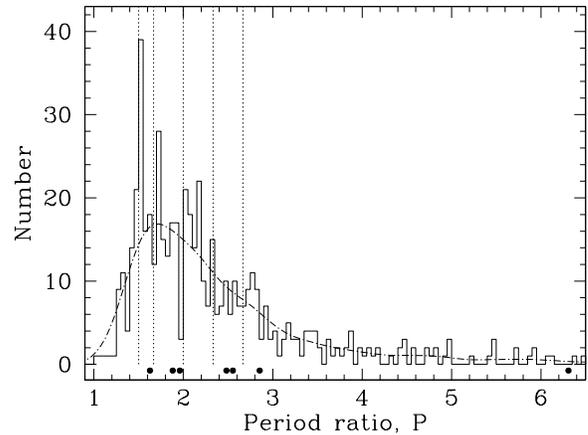}
\vskip-0.1truein\caption{The period ratio distribution in multiple planet systems discovered by {\it Kepler}.  (Data from~\cite{Fabrycky:2014}.)
The dot-dashed curve is a smoothed version of the histogram (smoothed with a Gaussian kernel).  The black points indicate solar system values. The vertical dotted lines indicate locations of low order resonant values (3/2, 5/3, 2/1,7/3, 8/3).}
\label{fig:figP}
\end{figure}

We also define the period ratio for nearest-neighbor planets, $\P_i=T_{i+1}/T_i$.  
In Figure~\ref{fig:figP}, we plot the distribution of $\P_i$ for the ensemble of 373 multiple planet systems discovered by {\it Kepler}; there are 566 period ratios in this data.   It is a broad distribution, with paucity of period ratios close to 1, a peak near 1.6, and a long tail of large values.  There is also interesting fine structure within this broad distribution, particularly a trough-peak feature near low order resonant period ratios, such as 3/2 and 2/1, that has been discussed in several recent papers~\citep{Lithwick:2012,Batygin:2013,Petrovich:2013,Fabrycky:2014}.  In the present work, we attempt to understand the overall distribution of $\P$. % and how it may inform inferences of planetary masses.  

%%%***
%We can ask about the selection biases and incompleteness of the apparent distribution of $\P$.  If a higher sensitivity survey, analogous to the {\it Kepler} survey, were to be carried out, we might expect that smaller planets would be discovered in greater numbers, over the same range of orbital periods (limited by the length and cadence of the survey).  It is not immediately obvious how this would affect the distribution of $\P$.  \cite{Steffen:2013} has compared the distribution of $\P$ of the high multiplicity and the low multiplicity systems within the {\it Kepler} sample and concluded that the period ratio distribution of {\it Kepler}'s harvest of multi-planet systems is a fair sample of the intrinsic distribution, at least for $\P\lesssim5$ or 6.   In the analysis below, we will assume that the observed distribution of $\P$ for the {\it Kepler} multi-planet systems is a proxy for its intrinsic distribution;  corrections to this distribution with future advances can be easily incorporated in the approach we describe below to derive the planet mass distribution function from the distribution function of $\P$.  
%%%***

%%%%%%%%%%%%%%%%%%%%%%%%%%%%%%%%%%%%%%%
%%%%%% Figure-P %%%%%%%%%%%%%%%%%%%%%%%%%%
%%%%%%%%%%%%%%%%%%%%%%%%%%%%%%%%%%%%%%%%%%
%%%%%%%%%%%%%%%%%%%%%%%%%%%%%%%%%%%%%%%%%%

Examining Fig.~\ref{fig:figP}, it is not difficult to be persuaded that the steep drop in the $\P$ distribution  from near $\P\approx 1.5$ to $\P\approx 1.3$ and the paucity of systems with $\P$ close to unity is likely owed to the instability of very closely spaced orbits, due to mutual planetary perturbations: larger planet masses require larger orbital spacings for long term dynamical stability; higher planet multiplicity and higher orbital eccentricities also would tend to require larger orbital spacing for stability (with the exception of librating resonant orbits).  Therefore, if the orbital spacings in multi-planet systems are related to their long term dynamical stability, the planet masses must be related to the orbital spacings.  By use of Kepler's third law, the dimensionless orbital spacing, ${\cal D}_i$, is related to the period ratio of adjacent planets, 
\begin{equation}
{\cal D}_i \equiv 2{a_{i+1}-a_i\over a_{i+1}+a_i} = 2{\P_i^{2\over3}-1\over \P_i^{2\over3}+1}.
\label{e:Delta}
\end{equation}
Note that $0<\D_i<2$.  

\begin{figure}[t]
\centering
\vskip-0.25truein\includegraphics[scale=0.35,angle=270]{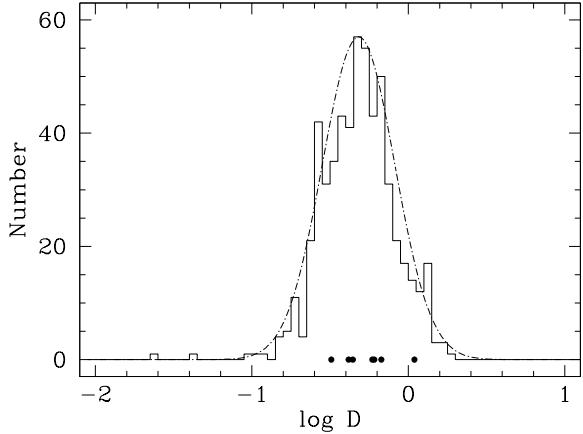}
\vskip-0.1truein\caption{The distribution of the orbital spacing, ${\D}$ (Eq.~\ref{e:Delta}), of adjacent planets in multiple planet systems discovered by {\it Kepler}.  The dot-dashed curve is the best-fit Gaussian function.  The black points indicate solar system values. }
\label{f:figD}
\end{figure}

The observed distribution of $\log\D$ for {\it Kepler} planets is shown in Figure~\ref{f:figD}.  We find that a Gaussian function, with mean $\bar x_\D\equiv\langle\log{\cal D}\rangle=-0.318$ and standard deviation $\sigma_{\cal D}=0.231$, fits the data fairly well; a $\chi^2$ test gives $p$--values $>0.1$.  Of course, the Gaussian must be truncated at a maximum value, $\log\D=\log 2$.  Therefore, formally, the best-fit probability density function (PDF) for $X=\log\D$ is expressed as
\begin{equation}
F_{\D}(x) = \Bigg\{ \begin{array}{ll}
{1\over{\sqrt{2\pi}\sigma_\D\Phi_\D}}\exp[-{(x-\bar x_{\D})^2\over 2\sigma_{\D}^2}] 
   &\mbox{ if $x<\log 2$}\\
0 & \mbox {otherwise,}
\end{array}
\label{e:FD}\end{equation}
where 
\begin{equation}
\Phi_{\D} = \Phi({\log 2-\bar x_\D\over \sigma_\D}),
\label{e:PhiD}\end{equation}
and $\Phi(\cdot)$ is the cumulative distribution function of the standard normal distribution.  For the computed values of the mean and standard deviation, we find $\Phi_D\simeq 0.996$. Therefore we incur only a very small error in approximating $F_\D(x)$ as an untruncated Gaussian function.

%%%%%%%%%%%%%%%%%%%%%%%%%%%%%%%%%%%%%%%%%%
%%%%%% Figure-D %%%%%%%%%%%%%%%%%%%%%%%%%%
%%%%%%%%%%%%%%%%%%%%%%%%%%%%%%%%%%%%%%%%%%
%%%%%%%%%%%%%%%%%%%%%%%%%%%%%%%%%%%%%%%%%%

In other words, the PDF of the dimensionless orbital spacing, $\D$, is a nearly log-normal distribution.  A log-normal is a skewed distribution; the mean, median and mode of $\D$ are 0.554, 0.481 and 0.362, 
%$10^{\bar x_\D+{1\over2}\sigma_D^2\ln10}=0.554$,  
%$10^{\bar x_\D}=0.481$, 
%$10^{\bar x_\D-\sigma_D^2\ln10}=0.362$, 
respectively.  Note that a log-normal distribution can resemble a power-law distribution over a fairly wide range of the parameter away from the peak.

We remark that the dimensionless orbital separations of solar system planets (indicated by the black dots in Fig.~\ref{f:figD}) are not dissimilar to those of the {\it Kepler} planets.

\section{Dynamical stability}

%%%%%%%%%%%%%%%%%%%%%%%%%%%%%%%%%%%%%%%%%%
%%%%%% Table-solsys %%%%%%%%%%%%%%%%%%%%%%%%%%
\begin{table}[b]
\begin{center}
\caption{Solar system planets}
\begin{tabular}{l c c c c c c}
\hline\\
Planet pair & $\tilde\mu$ & $\P$ & $\gamma$ & $K$   \\ \\ \hline
Venus-Mercury & 2.61E-06 & 2.55 & 0.678E-01 & 63.4  \\
Earth-Venus & 5.45E-06 & 1.63 & 0.815 & 26.3  \\
Mars-Earth & 3.33E-06 & 1.88& 0.107 & 40.1  \\
Jupiter-Mars & 9.55E-04 & 6.31 & 0.338E-03 & 16.0  \\
Saturn-Jupiter & 1.24E-03 &2.48 & 0.299 & 7.90  \\
Uranus-Saturn & 3.29E-04 & 2.85 & 0.153 & 14.0  \\
Neptune-Uranus & 9.52E-05 &1.96 & 0.848 & 14.0 \\ \hline
\end{tabular}
%\tablecomments{}
\end{center}
\end{table}
%%%%%%%%%%%%%%%%%%%%%%%%%%%%%%%%%%%%%%%%%%
%%%%%%%%%%%%%%%%%%%%%%%%%%%%%%%%%%%%%%%%%%

Let's consider the dynamical stability of a pair of adjacent planets in the simplest case, that of a two-planet system.  For nearly co-planar and initially circular orbits, the smallest orbital spacing that is dynamically stable is given by Hill's criterion:
\begin{equation}
\D = K \Big({\tilde\mu\over3}\Big)^{1\over3},
\quad \hbox{with}~ K=2\sqrt{3}. %\simeq 3.46. 
\label{e:khill}
\end{equation}
Note that this stability criterion is insensitive to how the total planet mass is partitioned between the two planets.  It is independent of the actual distance of the planets from the host star.

For systems with more than two planets, there is no known analytic criterion for dynamical stability, but we expect that in close-packed systems of $N\ge3$ planets, orbital spacings must exceed those required by  Hill's criterion, i.e., $K$ must exceed $2\sqrt{3}$.  
We can  look to numerical studies of dynamical stability of planetary systems for insights.  Several such investigations have been published, many focussed on particular systems, but a few on the broader theoretical question of the dynamical lifetimes of multiple planet systems as a function of planet masses and orbital spacings.  Numerical results for nearly coplanar, low eccentricity multi-planet systems (of $3\le N\le20$ equal-mass planets with $10^{-9}\le\mu_i\le10^{-5}$) show that dynamical stability times %(usually defined as the timescale for orbit crossings to occur) 
in excess of $\sim10^8~T_1$ require that adjacent planet pairs must have $K\gtrsim8$ and that this lower limit on $K$ is only weakly dependent on planet mass and planet multiplicity~\citep{Chambers:1996, Smith:2009}.  For higher planet masses $(10^{-3.4}\lesssim\mu_i\lesssim10^{-2.4})$, the critical $K$ is somewhat smaller, $K\gtrsim5$~\citep{Marzari:2002,Chatterjee:2008}.   The notation ``$\Delta$", ``$\beta$'' and $K$ has been used by various authors for the same parameter; here we have adopted $K$. This parameter, which is the orbital separation in units of the mutual Hill radius, is often called ``dynamical separation". 

 An important insight from these studies is that the chaotic nature of multiple planet systems necessitates a statistical description of the relationship between planet masses, orbital separations and the system's dynamical stability time, i.e., that $K$ depends sensitively on initial conditions and is better described as a random variate.  The distribution of $K$ depends upon the  planetary architecture.  For the singular but well-studied example of the solar system, investigations of its long term dynamics have concluded that it is marginally stable on timescales comparable to its age~\citep{Laskar:1996,Hayes:2008,Hayes:2010,Lithwick:2014,Batygin:2015}.  Taking adjacent pairs of solar system planets, we find that $K$ ranges from $\sim8$ (for Jupiter-Saturn) to $\sim63$ (for Mercury-Venus), with a mean value of 26 (see Table~1).  
 These values are larger than required by the two-planet Hill's stability criterion, and reflect the effects of non-trivial orbital eccentricities, mutual inclinations, unequal planet masses and planet multiplicity.
With this as motivation, we adopt a heuristic criterion for the long term stability of systems with more than two planets: a straightforward generalisation of Eq.~\ref{e:khill} in which we treat $K$ as an independent random variate.   We will denote by $P_K(\cdot)$ the PDF of $K$, and denote by $F_K(\cdot)$ the PDF of $\log K$. 

\section{Analytical estimates of the distribution of total mass of adjacent planet pairs}

The following rearrangement of Eq.~\ref{e:khill} will be useful for our analysis, 
\begin{equation}
\log\tilde\mu = 3(\log\D-\log K) + \log 3.
\label{e:lmutilde}
\end{equation}
We can then estimate the probability density function $F_{\tilde\mu}(\cdot)$ of $\log\tilde\mu$, as a convolution of the PDFs of $\log\D$ and of $\log K$,
\begin{equation}
F_{\tilde\mu}(x) = \begin{array}{ll}
& \int_{-\infty}^{\infty}d{x_\D}\int_{-\infty}^{\infty} dx_K  F_K(x_K) F_{\D}({x_\D}) \\
&\qquad\qquad\times\delta(x - 3x_\D+3x_K-\log3),
\end{array}
\label{e:Pmut}
\end{equation}
where $\delta(x)$ is the Dirac delta function.  
One challenge we face is that the published numerical studies have not reported the PDF of $K$, nor have they explored the large parameter space of systems containing unequal mass planets.  This requires systematic numerical investigation, which is feasible with modern computers, but has yet to be undertaken.  In the absence of such knowledge, we will adopt some plausible ansatzs for the PDF of $K$.  

As a simple illustration, let's first consider $F_K(x_K)=\delta(x_K-\log K_*)$.  Then, with the best-fit Gaussian function for the PDF of $\log\D$~(Eq.~\ref{e:FD}), it is straightforward to determine that the distribution of $\log\tilde\mu$ is also Gaussian, with mean $\langle\log{\tilde\mu}\rangle=-0.48-3\log K_*$, and standard deviation $\sigma_{\tilde\mu}=0.693$.  If we set $K_*=2\sqrt3$, the minimum value needed for dynamical stability of two planets, then $\log\tilde\mu$ has a Gaussian distribution of mean $\langle\log\tilde\mu\rangle=-2.09$ and standard deviation 0.693.  The corresponding distribution of $\tilde\mu$ is log-normal, with median value $10^{-2.09}$ and mode $10^{-3.20}$. Because this describes the maximal $\tilde\mu$ that is dynamically stable, we  can conclude that the most common planet mass does not exceed $10^{-3.20}m_*$, or about two-thirds Jupiter mass for solar mass stars.

For a more realistic estimate, let's consider a broad distribution of $K$ values. 
We are inspired to consider a Gaussian distribution for its simplicity, because it follows from Eq.~\ref{e:lmutilde} that $\log\tilde\mu$ has a Gaussian distribution if $\log\D$ and $\log K$ are both Gaussian variates.  We are also motivated by the $K$ values in the solar system (Table~1): $\log K$ is in the range 0.9 to 1.8, with mean $\langle\log K\rangle=1.32$ and standard deviation $\sigma_K = 0.31$.   Although the sample is small, its $\log K$ distribution is not inconsistent with a normal distribution. 
For $F_K(x_K)$, we therefore adopt a Gaussian function\footnote{Formally, the PDF of $\log K$ must vanish for $K < 2\sqrt3$, to satisfy the Hill criterion (Eq.~\ref{e:khill}) for the limiting case of two planet systems.  This means that we should adopt a truncated Gaussian PDF, analgous to the case of the PDF of $\log D$~(Eq.\ref{e:FD}).  For the parameters of interest here, the normalization factor is nearly unity, so we neglect the truncation.}, with the mean and standard deviation chosen to match the values found for the solar system.  
Then, it is straightforward to compute that the resulting Gaussian PDF for $\log\tilde\mu$ has mean $\langle\log\tilde\mu\rangle=-4.44$ and standard deviation $\sigma_{\tilde\mu}=1.16$.  The median value of $\tilde\mu$ in this case is $10^{-4.44}$; the mode is  $\sim10^{-7.5}$, about three orders of magnitude smaller than the upper bound derived above with $K_*=2\sqrt3$.

It is evident that dynamical stability implies that the PDF of $\log{\tilde\mu}$ has small probability density at both small and large values and a peak at an intermediate value.   This shape is inherited from the nearly Gaussian distribution of $\log\D$, which in turn is derived from the observed distribution of period ratios.  In particular, the drop-off at small masses is inherited  from the steep drop-off in the number of observed systems with period ratios smaller than $\sim1.5$.   

We remark that the mean value of $\log\tilde\mu$ decreases with increasing mean $K$.  However, the peaked shape of the PDF of $\log\tilde\mu$ is not severely dependent on the particular functional form of the $K$ distribution that we adopted.   For example, a flat, uniform random distribution of $K$ in a range $K_{min}\ge2\sqrt3$ to $K_{max}=95$  (slightly wider than the range found in the solar system) also yields a peaked distribution of $\log\tilde\mu$.  The skewness of the $\tilde\mu$ distribution does depend on the dispersion of the $K$ distribution, and, consequently, the most probable value of $\tilde\mu$ depends upon this dispersion as well.  Our choice of the PDF of $K$ is of course motivated by the solar system and could be considered biased; we discuss this point further in Section~6.

\section{Numerical estimates of the planet mass distribution function}

To determine the distribution of individual planet masses requires additional considerations: are the masses of adjacent planet pairs correlated or are they independent?  

Let $P_{\mu}(x)$ be the probability density function of $\mu_i$ (planet mass as fraction of stellar mass).  
Let's consider the limiting case in which adjacent pairs of planets have mass ratio which is a fixed constant, $\gamma_i=\gamma_*$.  Then the PDF of $\mu_i$ is straightforwardly derived from that of $\tilde\mu$,
\begin{equation}
P_{\mu}(x) = {1\over2}[ (1+\gamma_*) P_{{\tilde\mu}}((1+\gamma_*)x)+ (1+\gamma_*^{-1})P_{{\tilde\mu}}((1+\gamma_*^{-1})x) ]. 
\label{e:Pmu}\end{equation}
On the other extreme, if the masses of adjacent planet pairs are considered independent, then
we have the following relationship between $P_\mu$ and $P_{\tilde\mu}$,
\begin{equation}
P_{\tilde\mu}(x) = \int_0^x dy~ P_{\mu}(y) P_{\mu}(x-y). 
\label{e:Pmutilde}\end{equation}
We will not attempt to solve this implicit equation for $P_{\mu}$, but we can note that, depending on its functional form, the mass ratio, $\gamma$, of planet pairs may or may not be independent of $\tilde\mu$, even if the masses of adjacent planets were independent.    

In reality, adjacent planet masses are likely to be neither perfectly correlated nor perfectly independent.  We therefore consider $\gamma$ to be a random variate, and for simplicity, we assume that it is independent of the total mass of the planet pairs. 
We consider two cases:  (i) a uniform PDF for $\gamma$, $P_\gamma(x) =1 $ or 
(ii) a PDF with peak at 1.  For the latter, we chose $P_\gamma$ to be a Gaussian with mean 1, standard deviation 0.3, truncated at 0 on the left and 1 on the right; in this distribution,  approximately half of all planet pairs have $0.8<\gamma\le1$). This distribution is similar to the assumption that neighboring planets tend to have similar masses, whereas the uniform  distribution allows any value of their mass ratio with equal probability.    %In adopting these distributions of $\gamma$, we have made the simplifying assumption that it is independent of the total mass of planet pairs.

We carry out the numerical calculation of the planet mass distribution function as follows.  Starting with the observational data of the period ratios, $\P_i$, of the {\it Kepler} planets, we first calculate $\D_i$. Then we calculate $\tilde\mu_i$ with the help of Eq.~\ref{e:lmutilde} and a random value of $K$ from its prescribed PDF; we adopt a Gaussian PDF of $\log K$ with mean 1.32 and standard deviation 0.31, as discussed in the previous section.  Next we compute the individual $\mu_i$'s with the help of Eq.~\ref{e:muitilde}--\ref{e:gamma} and a random value of $\gamma$ from its prescribed PDF.  We average over 1000 realizations of the random choices of $\log K$ and $\gamma$.
We take one additional step: to compute the individual planet masses, $m_i=\mu_i m_*$, we adopt the stellar host masses of the {\it Kepler} multiple-planet systems (obtained from estimates of the stellar surface gravity and stellar radius by~\cite{Batalha:2013}, as reported in \cite{Fabrycky:2014}).  

\begin{figure}[!]
\centering
\vskip-0.25truein\includegraphics[scale=0.35,angle=270]{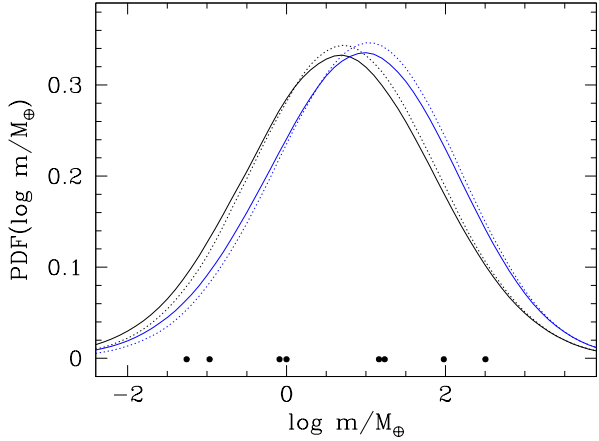}
\vskip-0.15truein\caption{The distribution of log-mass of {\it Kepler} planets, derived from their period ratios and a heuristic criterion for dynamical stability.  The continuous line curve and dot-dashed curve are the results obtained by assuming a uniform random and a half-Gaussian distribution, respectively, of the ratio of adjacent planet masses, $\gamma$ (Eq.~\ref{e:gamma}).  The black curves are based on the observed period ratios, while the blue curves are based on the de-biased distribution of period ratios.  The black points indicate the masses of solar system planets.  
}
\label{f:m1}
\end{figure}

The results are shown in Fig.~\ref{f:m1}, where we plot in continuous line the case of $\gamma$ uniform random distribution on $(0,1)$, and in dot-dashed line the case of $\gamma$ having a half-Gaussian PDF peaked at 1.  
We observe that (a) the estimated planet mass function is not very sensitive to the choice of $P_\gamma$, and (b) the PDF of the logarithm of planet mass does not increase monotonically as the mass decreases.  The distribution of $\log m/M_\oplus$ is found to be peaked, with mean $0.64(0.72)$ and standard deviation $1.21 (1.17)$ for $\gamma$ uniform random (half-Gaussian peaked at 1).

%%%%%%%%%%%%%%%%%%%%%%%%%%%%%%%%%%%%%%%
%%%%%% Figure-m1 %%%%%%%%%%%%%%%%%%%%%%%%%%
%%%%%%%%%%%%%%%%%%%%%%%%%%%%%%%%%%%%%%%

%%%***
The above analysis is based on the observed orbital period ratios of {\it Kepler} planets.
We can ask about the selection biases and incompleteness of the observed distribution of $\P$.  If a higher sensitivity survey, analogous to the {\it Kepler} survey, were to be carried out, we might expect that smaller planets would be discovered in greater numbers, over the same range of orbital periods (limited by the length and cadence of the survey).  It is not immediately obvious how this would affect the distribution of $\P$.  \cite{Steffen:2013} has compared the distribution of $\P$ of the high multiplicity and the low multiplicity systems within the {\it Kepler} sample and concluded that the period ratio distribution of {\it Kepler}'s harvest of multi-planet systems is a fair sample of the intrinsic distribution, at least for $\P\lesssim5$ or 6.  
\cite{Steffen:2015} have analyzed the incompleteness of the observed $\P$ distribution due to planets missed by the {\it Kepler} data reduction pipeline as well as due to geometric bias against detection of non-coplanar planets.  The authors assumed that the mutual inclinations of planetary orbits have a Rayleigh distribution with width parameter $\sigma=1.5$~degrees.  Their debiased $\P$ distribution is broadly similar to the observed distribution, but has higher probability densities for larger period ratios (see their Figure 4).  Using this debiased $\P$ distribution, we repeated our calculation of the planet mass distribution; the results are plotted in blue in Figure~\ref{f:m1}.  We find that the resulting distribution of $\log m/M_\oplus$ has mean $0.91(0.98)$ and standard deviation $1.18(1.14)$ for $\gamma$ uniform random (half-Gaussian peaked at 1).   Steffen (2015, personal communication) also provided us with a debiased $\P$ distribution based on a model Rayleigh distribution of the mutual inclinations with width parameter $\sigma=3^\circ$; the resulting planet mass distribution is insignificantly different than that for $\sigma=1.5$~degrees.

In summary, we find that the planet mass distribution derived from use of the heuristic dynamical stability criterion and the observed and debiased distribution of period ratios of {\it Kepler} planets is peaked in logarithm of planet mass, with a peak value of $\log m/M_\oplus$ of $0.6-1.0$, and standard deviation $1.1-1.2$.
 %%%***
     
\section{Comparison with other estimates}

All the planets discovered by {\it Kepler} have fairly well determined values of their planetary radii, with uncertainties of about 30\%~\citep{Silburt:2015}.  Several studies have proposed empirical relations between planet masses and planet radii.  \cite{Lissauer:2011} have proposed the following, based on the well-known properties of the solar system planets:
\begin{equation}
m/M_\oplus  = (R/R_\oplus)^\alpha,
\label{e:lissauer}\end{equation}
where $M_\oplus$ and $R_\oplus$ are the mass and radius of Earth, $\alpha=2.06$ if $R>R_\oplus$ and $\alpha=3$ if $R\le R_\oplus$.  \cite{Wu:2013} have proposed a slightly different relation, based on observational data of a subset of {\it Kepler} planets whose masses have also been determined observationally:
\begin{equation}
m/M_\oplus  = 3(R/R_\oplus).
\label{e:wu}\end{equation}
\cite{Weiss:2014} report a slightly different empirical mass-radius-density relation, based on an updated list of {\it Kepler} planets of radius $R<4R_\oplus$ whose masses have been measured; this can be expressed as follows:
\begin{equation}
m/M_\oplus= \Bigg\{ \begin{array}{ll}
%((2.43+3.39R/R_\oplus)/5.51)(R/R_\oplus)^3
(0.441+0.615R/R_\oplus)(R/R_\oplus)^3\ 
   \mbox{ if $R<1.5R_\oplus$},\\
2.69(R/R_\oplus)^{0.93}\ \mbox {if $1.5R_\oplus\le R<4R_\oplus$}.
\end{array}
\label{e:weiss}\end{equation}
%where we adopted $\rho_\oplus=5.51$~g~cm$^{-3}$ for the mean density of Earth.

Using each of these mass-radius relations and the data for planetary radii (reported in \cite{Fabrycky:2014}), we computed the masses of the 939 {\it Kepler} planets in multiple-planet systems, and then used Gaussian kernel density estimation to compute the PDF of the logarithm of masses.  
For $R\ge4R_\oplus$, we supplemented Eq.~\ref{e:weiss} with Eq.~\ref{e:wu}.

\begin{figure}
\centering
\includegraphics[scale=0.35,angle=270]{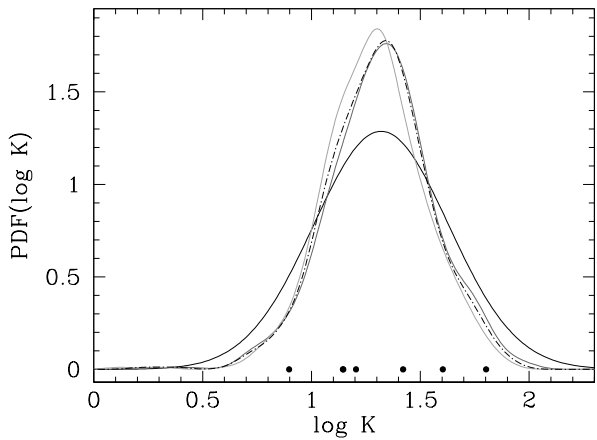}
\caption{Comparisons of the distribution of $\log K$, where $K$ is the orbital spacing in units of the mutual Hill radius.  The dark grey, light grey and dot-dashed curves are the results obtained from the mass-radius relations of~Eq.~\ref{e:lissauer}, Eq.~\ref{e:wu}, and Eq.~\ref{e:weiss}, respectively. The black continuous line curve is the Gaussian distribution that we adopted, with mean and standard deviation matching that of the solar system planets. The black points indicate solar system values. 
}
\label{f:figk}
\end{figure}

As an aside, we note that, having computed the planet masses by using mass-radius relations, we then computed $K$ values for the adjacent planet pairs.  From these values, we estimated the PDFs of $\log K$ using Gaussian kernel estimation; these are shown in Fig.~\ref{f:figk}.  We find that the three empirical mass-radius relations yield similarly peaked distributions of $\log K$, with mean values 1.32, 1.29 and 1.31, and standard deviations 0.24, 0.23 and 0.23, respectively.  The mean values are similar to that of the solar system planets; the standard deviations are somewhat smaller.   %Evidently, as measured by the distribution of $K$, the solar system is not an outlier compared to the {\it Kepler} systems. 
This provides support for the PDF of $\log K$ that we adopted by ansatz.

%%%%%%%%%%%%%%%%%%%%%%%%%%%%%%%%%%%%%%%
%%%%%% Figure-Fk %%%%%%%%%%%%%%%%%%%%%%%%%%
%%%%%%%%%%%%%%%%%%%%%%%%%%%%%%%%%%%%%%%

\begin{figure}
\centering
\includegraphics[scale=0.35,angle=270]{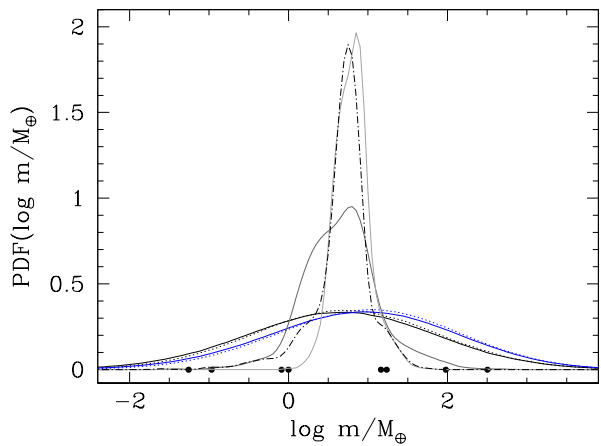}
\caption{Comparisons of PDFs of the logarithm of planet mass: The dark grey, light grey and dot-dashed curves are the results from the mass-radius relations of~Eq.~\ref{e:lissauer}, Eq.~\ref{e:wu} and Eq.~\ref{e:weiss}, respectively.   The black and blue continuous and dotted curves are our theoretical estimates (as in Fig.~\ref{f:m1}). 
}
\label{f:comparisons}
\end{figure}

The mass distributions obtained by use of the mass-radius relations (Eq.~\ref{e:lissauer},  Eq.~\ref{e:wu}, Eq.~\ref{e:weiss})  are shown in Fig.~\ref{f:comparisons}. 
We see that these yield strongly peaked PDFs of $\log m/M_\oplus$, with mean values $0.66$, $0.80$ and $0.68$, respectively. 
For comparison, we also plot our theoretical estimates based on dynamical stability. 
The PDFs derived independently from the mass-radius relations peak near similar values to our dynamical stability--based PDFs.  The latter have significantly larger dispersion, however.  The smaller dispersions of the former are not entirely surprising, as the mass-radius relations describe empirical best-fits and do not reflect the uncertainties and dispersion in the observational data.  \cite{Weiss:2014} note that the observational data have significant scatter about their empirical mass-radius relation, and that the scatter  is not merely due to observational errors but may reflect intrinsic compositional diversity of planets.

%%%%%%%%%%%%%%%%%%%%%%%%%%%%%%%%%%%%%%%
%%%%%% Figure-comparisons %%%%%%%%%%%%%%%%%
%%%%%%%%%%%%%%%%%%%%%%%%%%%%%%%%%%%%%%%

Overall, in comparison with the estimates based on mass-radius relations, the dynamical stability--based estimate has a shallower power law slope of the planet mass function at planet masses larger than about ten Earth-mass and a steeper slope for masses below about one Earth-mass.  %This is evident in Fig.~\ref{f:comparisons2} which displays, in a log-log plot, the PDFs of $m/M_\oplus$ for each of the PDFs of $\log m/M_\oplus$ displayed in Fig.~\ref{f:comparisons}.  
These discrepancies may reflect an overestimate of the width of the empirical distribution of $\log K$ adopted in our dynamical stability-based estimate, or they may reflect observational incompleteness (due to undetected small and/or long period planets) and the intrinsic scatter about the best-fit mass-radius relations.  Future studies can test these alternate hypotheses.

\section{Discussion and Summary}

A quantitative description of the architectures of planetary systems requires at least the following: the degree of orbital co-planarity, the distribution of their orbital periods and spacings, and the distribution of planetary masses.  The large number of planetary systems discovered by the {\it Kepler} mission allows a statistical assessment of these properties that have long eluded the theory of the formation and evolution of planetary systems.  Several studies of the {\it Kepler} data suggest that planetary systems are flat, to within a few degrees, similar to our own solar system
\citep{Lissauer:2011,Tremaine:2012,Fang:2012b,Fabrycky:2014}.  Orbital spacings have also been the subject of several studies~\citep{Lissauer:2011,Steffen:2013,Fabrycky:2014,Steffen:2015}, as have planet masses~\citep{Howard:2010,Wu:2013,Weiss:2014}.  In the present work, we studied the distribution of orbital spacings and used dynamical stability to estimate planet masses in a statistical way. 

Regarding their orbital spacings, we found that the {\it Kepler} planets' dimensionless orbital spacings, $\D$ (Eq.~\ref{e:Delta}), have a nearly log-normal distribution~(Fig.~\ref{f:figD}). How can we understand this distribution of $\D$?  In qualitative terms, a log-normal distribution is generated in a multiplicative random process, that is, when a positive definite variable $v$ suffers random increments in proportion to its value, $\delta v \propto v$.  If the successive increments are independent and large in number, then by the central limit theorem $\log v$ will be approximately normally distributed.  We conjecture that the log-normal distribution of $\D$ arises in the late stages of the dynamical evolution of planetary systems when the planetary architecture is shaped (or re-shaped) by secular chaos~\citep{Lithwick:2014}, or possibly it arises in an earlier stage when a small number of planets emerge from their natal protoplanetary disk but still embedded in a leftover disk of a large number of planetesimals.  As a consequence of mergers or ejections of planetesimals and/or planets, the surviving planets undergo a random walk of their orbits; unstable configurations are steadily winnowed.  Our understanding of this evolution is still at an early stage~\citep{Rein:2012,Hansen:2013,Minton:2014,Hands:2014,Chatterjee:2015}. In future studies, it would be very useful to examine quantitatively how (or if) secular chaos and/or planetesimal-aided orbital migration lead to a log-normal distribution of orbital spacings.  

Regarding the masses of planets, we reason that it must be possible to deduce planetary masses from the observational data of orbital period ratios if orbital spacings are determined by dynamical stability.  However, there is no direct way to do so, therefore we made several simplifying assumptions and ansatzs.  First, we used the two--planet Hill's stability criterion to derive an upper bound for the most common planet mass.  Then, we generalized Hill's criterion in a statistical way for the stability of multi-planet systems~(Eq.~\ref{e:lmutilde}) to compute the planet mass function.  We assumed that the dynamical separation (i.e., the orbital separation in units of the mutual Hill radius) and the mass ratio of adjacent planets are both independent random variates. We adopted plausible distribution functions for these two parameters, based on our understanding of solar system dynamics.  These empirical distributions and the assumed independence of the variables are admittedly major simplifications.  These simplifications can be relaxed in a future study by determining the joint probability density function of $K$, $\tilde\mu$ and $\gamma$ by means of large scale numerical simulations of the dynamical stability of multiple planet systems.

The multi-planet systems discovered by {\it Kepler} have been described as being rather unlike the Solar system, because their orbital periods are $\sim10$~days and their masses are on the order of a few Earth masses; such planets are absent in the solar system.  However, the dimensionless orbital spacings and the dynamical separations of solar system planets are not dissimilar to those of the {\it Kepler} systems~(Figs.~\ref{f:figD} and~\ref{f:figk}), when we compute the latter independently from the observational data for planetary radii by using mass-radius relations.  By these scaled measures of ``planetary system architecture'', the solar system does not appear to be an outlier.  

A deeper look at the dynamical separations in the solar system shows that the terrestrial planets, Mercury--Mars, have $K$ values significantly larger than those of the outer planets, Jupiter--Neptune (see Table~1).  This was pointed out in \cite{Ito:2002}. It is also notable that the $K$ values for the giant planets (Jupiter--Neptune) are closer to the numerically determined minimum value, $K_*\approx5-8$, that is necessary for the dynamical stability of $N\ge3$ equal mass planetary systems for timespans of the age of the solar system, whereas the $K$ values for the terrestrial planets are significantly higher.  We conjecture, following \cite{Ito:2002}, that this dichotomy is owed to the property that the subsystem of the four terrestrial planets is subject to long term dynamical excitation by the giant planets.  This is supported by the numerical experiments reported in~\cite{Hayes:2010} who found that the terrestrial planets exhibit a much lower level of long term chaos if the gravitational perturbations of the giant planets were absent.  It is also possible that $K$ is related to ``dynamical age'', i.e., the age of the system measured in units of the orbital period of the innermost planet.  We observe that in the solar system, the terrestrial planets' dynamical age is 1 to 2 orders of magnitude larger than that of the outer planets (since the orbital period of Jupiter is about 50 times that of Mercury).
If so, then solar system--like planetary architectures may be better modelled with a bimodal PDF of $K$. For example, we could consider a PDF of $\log K$ consisting of an equally-weighted sum of two Gaussian functions having mean $\langle \log K\rangle$ of 1.06 and 1.61, respectively, and each having standard deviation 0.2; solar system $K$ values of the giant planets and terrestrial planets, respectively, are roughly consistent with these parameters.  With this choice of bimodal PDF of $\log K$, we repeated the numerical calculations described in Section 5 to compute the associated planet mass function.  The resulting distribution of $\log m$ is bimodal. The two peaks are near $\log m/M_\oplus$ values of $-0.2$ and $1.4$; the overall mean value of $\log m/M_\oplus$ is $0.60 (0.67)$ and the standard deviation is $1.41 (1.37)$ for $\gamma$ uniform random (half-gaussian).  The two peaks are of only slightly differing heights, and this PDF can also be described as approximately a plateau in the range $-0.2$ to $1.4$.  We do not belabor the results of this numerical experiment, however, because at this level of detail, we must also pay attention to possible correlations amongst the parameters, $K$, $\tilde\mu$, $\gamma$ and orbital periods.   A comprehensive numerical study of the dynamical stability of multiple-planet systems can provide an improved estimate of the joint probability distribution of these parameters.  %which may provide more refined diagnostics of solar system--like planetary architectures.  

In any study of the mass distribution of planets, the question arises of the definition of ``planet'', a question that has been debated in both public and scientific forums.   At the upper end of the planet mass range, the literature makes a distinction between ``giant'' planets and ``brown dwarfs'' near a mass of about $13M_J(\approx4134M_\oplus)$.   In the solar system,  ``planet'' masses range from $0.055M_\oplus$ to $318M_\oplus$, and, at the lower end of the mass range, the literature also recognizes ``dwarf planets'', the most massive being $\sim0.002M_\oplus$.  The results of our dynamical stability--based estimate of the mass function are most pertinent for the mass range spanning a few percent of an Earth-mass to a few hundred Earth-masses.  Although these results can be smoothly extended beyond these limits, it is likely that different physical processes shape the mass function near these upper and lower bounds.

Our results and conclusions are summarized as follows. 

\begin{enumerate}

\item
The observed period ratios of adjacent planet pairs in multiple-planet systems discovered by {\it Kepler} indicate that the distribution of the dimensionless orbital separation (Eq.~\ref{e:Delta}) is approximately a log-normal function.  

\item
The minimum dynamical separation (the orbital separation of adjacent planets in units of their mutual Hill radius, Eq.~\ref{e:khill}), $K=2\sqrt3$, necessary for long term dynamical stability of two--planet systems implies that the most common planet mass does not exceed $10^{-3.2}m_*$.  For solar mass stars, this is about two-thirds the mass of Jupiter.

\item
For plausible distributions of $K$ and of the mass ratio, $\gamma$ (Eq.~\ref{e:gamma}), of adjacent planet pairs, our theoretical estimate of the planet mass distribution function is peaked in $\log m$.  It is only weakly sensitive to the distribution of $\gamma$.  We estimate that the most probable value of $\log m/M_\oplus$ is $\sim(0.6-1.0)$, and the standard deviation of the distribution of $\log m/M_\oplus$ is about 1.2. 

\item
The planet mass distribution computed independently from the observational data on planetary radii (by use of empirical mass-radius relations) is also peaked in $\log m$ at similar peak values,  
but has smaller dispersions.  
These discrepancies may reflect an overestimate of the width of the distribution of $\log K$ adopted in our theoretical estimate, or they may reflect observational incompleteness of the measured masses at low and high planet masses and the intrinsic scatter of the masses about the best-fit mass-radius relations. Future studies can test these alternate hypotheses.

\item
In deriving the dynamical stability-based estimates, we assumed that $K$ and $\gamma$ are independent random variates. We adopted PDFs of $K$ and of $\gamma$ that are plausible, but arguably have a ``solar system bias''.  
A systematic numerical study of the dynamical stability of multi-planet systems, over a wide range of planet masses and orbital periods is needed to improve the theoretically expected distributions of planetary system parameters and corelations amongst them.   Such a study requires significant computational effort, but is feasible with modern computers.  This would enable an improved estimate of the planet mass function from observational data of orbital periods alone, which are readily measured in almost all observational methods currently employed for the detection of extrasolar planets.

\end{enumerate}

\acknowledgements
I thank Scott Tremaine, David Frenkel and Randy Jokipii for discussions.  I also thank an anonymous reviewer for helpful comments.  This research was supported by NSF (grant AST-1312498) and NASA (grant NNX14AG93G), and made use of the NASA Astrophysics Data System Bibliographic Services and the NASA Exoplanet Archive.  

\bibliographystyle{apj}

\end{document}